\documentclass[11pt,biometrika]{article}

\usepackage{mathtools}
\usepackage{amsmath}
\usepackage{amssymb}
\usepackage{natbib}
\usepackage{theorem}
\usepackage[T1]{fontenc}
\usepackage[latin1]{inputenc}
\usepackage{moreverb}
\usepackage{shortvrb}
\usepackage{textcomp}
\usepackage[english]{babel}
\usepackage{mathtools,xparse}
\usepackage{color}

\usepackage{authblk} 
 
\DeclarePairedDelimiter{\norm}{\lVert}{\rVert}
\DeclarePairedDelimiter\abs{\lvert}{\rvert}%

\NewDocumentCommand{\normL}{ s O{} m }{%
  \IfBooleanTF{#1}{\norm*{#3}}{\norm[#2]{#3}}_{L_2(\Omega)}%
}

\makeatletter
\let\oldabs\abs
\def\abs{\@ifstar{\oldabs}{\oldabs*}}
\let\oldnorm\norm
\def\norm{\@ifstar{\oldnorm}{\oldnorm*}}
\makeatother

\def\n{\noindent}

\newtheorem{prop}{Proposition}
\newtheorem{theorem}{Theorem}

\newtheorem{corollary}{Corollary}
\newtheorem{remark}{Remark}

\title{{\bf Non-Smooth Backfitting for Excess Risk Additive Regression Model with Two Survival Time-Scales}}

\author[1]{Munir Hiabu	\thanks{munir.hiabu@sydney.edu.au }}
\author[2]{Jens P. Nielsen\thanks{Jens.Nielsen.1@city.ac.uk}}
\author[3]{Thomas H. Scheike \thanks{thsc@sund.ku.dk (Corresponding author)}}
\affil[1]{School of Mathematics and Statistics, University of Sydney, Camperdown NSW 2006, Australia}
\affil[2]{Cass Business School, City, University of London, 106 Bunhill Row, London, EC1Y 8TZ, United Kingdom}
\affil[3]{Department of Public Health, University of Copenhagen, {\O}ster Farimagsgade 5B,
1014 Copenhagen K, Denmark}

\begin{document}
\maketitle

\begin{abstract}

We present a new backfitting algorithm estimating the complex structured non-parametric survival model of Scheike (2001) without having to use smoothing. The considered model is a non-parametric survival model with two time-scales that are  equivalent up to a constant
that varies over the subjects. Covariate effects are
modelled linearly on each time scale by additive Aalen models. 
Estimators of the cumulative intensities on the two
time-scales are suggested by solving local estimating equations 
jointly on the two time-scales. We are able to estimate the cumulative intensities solving backfitting
estimating equations without using smoothing methods and we
provide large sample properties and simultaneous confidence bands. 
The model is applied to data on myocardial infarction providing a separation of  the two effects stemming  from
 time since diagnosis and  age.

\n Key words: Aalen model, counting process,
 disability model, illness-death model, 
 generalized additive models, 
 multiple time-scales, non-parametric estimation, 
 varying-coefficient models.

\end{abstract}

\section{Introduction}
In many bio-medical applications in survival analysis it
is of interest and needed to  use multiple time-scales. A 
medical study will often have a follow-up time 
(for example time since diagnosis) for patients of different ages, and
here both time-scales will contain important but different information
about how the risk of, for example, dying is changing. 
We therefore consider the situation with two time-scales that 
are equivalent up to a constant for each
individual, such as for example follow-up time and age. 
One may see this as arising from the 
the illness-death model, or the disability model,
where the additional time-scale may be duration in the illness
state of the model; see \cite{Keiding1991} for a general discussion
of these models. 
There is rather limited work on how to deal with
multiple time-scales in a biomedical context, see for example 
\cite{Oakes1995,Iacobelli2013}  and \cite{Duchesne2000} and references therein.
We present a non-parametric regression approach with two time-scales where
each time-scale contribute additively to the mortality. 
The regression setting models the 
effect of covariates by additive Aalen models 
on each time-scale \citep{aale:1989,huff:mcke:1991,abgk-book,ms06}.
This allows covariates to have
effects that vary on two different time-scales.
In a motivating example we consider patients that experience
myocardial infarction, and aim at predicting the
intensity considering the two time-scales age and time since myocardial infarction. 
As a consequence, we can make survival predictions for patients given their age
at diagnosis.  This model was considered previously by \cite{Scheike2001} where estimation
was based on smoothing for one of the time-scales. A study closely related to ours is  \cite{Kauermann2006} who studied the two most common time scales: age and duration. The underlying technical setting of \cite{Kauermann2006} was a multiplicative hazard model without covariates that is estimated via splines. In contrast our approach is an additive hazard model including covariates and estimating without smoothing.  Alternative smoothing methodologies to multiplicative hazard estimation includes  \cite{Linton:etal:03,huang:00,hastietib86, Lin:etal:16}.
None of the known multiplicative hazard approaches including the ones mentioned above are able to estimate without smoothing, include time varying covariate-effects, or are able to provide simultaneous confidence bands as the additive approach of this paper does provide. We do know that smoothing improves efficiencies of cumulatively estimated quantities, see \cite{Guillen:etal:07} for the simplest possible case. However, smoothing is also a complexity and experts applying survival analysis have developed a practical way of smoothing by eye the underlying rough non-parametric estimators of \cite{Kaplan:Meier:58, Nelson:72}.  The advantage of providing estimators without smoothing is that there can be no confusion from the complicated process of picking the smoothing procedure first and the amount of smoothing after that. Even if a smoothing approach is eventually used, then the smoothing free procedure would always count as a benchmark approach to check whether something went wrong during the smoothing. Our backfitting approach is different from standard backfitting in regression, see for example the smooth additive backtiffing approach of \cite{Mammen:etal:99}, where data is projected down via a smoothing kernel onto an additive subspace. In the backfitting approach of this paper, the non-parametric dynamics is only taking place in the two time directions, and the end result is therefore closer to the classical approach of \cite{Nelson:72} with a non-smooth estimator of the dynamics in the one-dimensional time axis. What is obtained through Aalen's additive hazard regression model on two time axis is that the dynamics of the two time effects are adjusted for covariaties in a way that keep the one-dimensional structure of the non-parametric dynamics. 
The expert user of survival methodology can therefore use the well developed intuition from looking at Nelson-Aalen estimators and Kaplan-Meier estimators when interpreting the empirical results based on the new methodology of this paper.
Another  advantage of estimating directly the cumulative hazards is that we are able to obtain a simple 
uniform asymptotic description of our estimators. We are thus
able to construct confidence bands and intervals, that are 
based on bootstrapping the underlying martingales.

The paper is organised as follows.
Section 2 presents the model via counting processes.
Section 3 gives some least squares based local estimating equations that are
solved to give simple explicit estimators of the
non-parametric
effects of the model. Based on these explicit estimators we are
able to derive asymptotic results and provide the estimators
with asymptotic standard errors. 
Sections 4-6 discusses how to solve the equations and compute the estimator
practically and how deal with identifiability issues. 
Section 7 shows how the large sample properties may be derived and in Section 8
we construct confidence bands.
Section 9 demonstrates the finite sample properties supporting 
Section 10  where we use our proposed methods in a worked example. 
Finally, Section 10 discusses some possible extensions.

\section{Aalen's Additive Hazard Model for Two Time-Scales} 

Let $N_i(t) \; \; i=1,...,n$ be $n$ independent counting 
processes
that do not have common jumps and are adapted to a filtration that 
satisfy the usual conditions \citep{abgk-book}.
We assume that the counting processes have intensities given by
\begin{align}\label{model}
\lambda_i(t) &= \sum_{j=1}^p X_{ij}(t)\alpha_j (t) +   \sum_{k=1}^qZ_{ik}(t)\beta_k( t+ a_i) \notag
\\&=X_i(t)\alpha (t) +  Z_i(t) \beta( t+ a_i), \quad (0\leq t \leq t_{max}),
\end{align}
where $\alpha=(\alpha_1, \dots, \alpha_p)$ and $\beta=(\beta_1,\dots,\beta_q)$ are tupels of one dimensional deterministic functions,
$X_i^T(t)  \in \Re^p$ and  $Z_i^T(t) \in \Re^q$  are  predictable cadlag covariate vectors with $X(t)$ and  $Z(t)$ having almost surely full rank, and $a_i$ is a
real-valued random variable observed at time $t=0$. 
If $Z_i(t)=0$ for all $t$, $a_i$ does not need to be observed. 


The model is the sum of two Additive Alalen Models running on two different time scales, see also Scheike(2001).
The two time-scales  are $t$ and $a=t+a_i \in [a_0,a_{max}]$ where the latter
time-scale is specific to each individual and  $a_0$ is some lower-limit that depends on the 
observed range of the second time-scale. 
Note, that no indicator variables are introduced but are absorbed
in the covariates. 
In the
illness-death model, say, $t$ might be time since diagnosis
(duration) 
among subjects that have entered the illness stage of the model
and 
$a_i$ could be the age when the transition 
to the illness stage occurred, such that $t+a_i$ is the
age  of the subject. 

After introducing some notation we present an 
estimation procedure that leads to explicit estimators 
of $A(t)=\int_0^t \alpha(s) ds=(\int_0^t \alpha_1(s) ds, \dots, \int_0^t \alpha_p(s) ds)^T$ 
and
$B(a)= \int_{a_0}^a \beta(u) du=(\int_{a_0}^a \beta_1(u) du, \dots, \int_{a_0}^a \beta_q(u) du)^T$.
The cumulative effects have the advantage compared to
$\alpha(s)$ and $\beta(a)$ that they may be used
for inferential purposes since a more satisfactory simultaneous
convergence can be established for these processes. 
We derive the asymptotic distribution
for these estimators and a bootstrapping procedure quantifying the estimation uncertainty. Based on the
cumulative intensity $A(t)$ one may estimate the intensity
$\alpha(t)$ by smoothing techniques.

\noindent {\bf 2.1\ \ Notation}
Let $\Lambda_i(t) = \int_0^t \lambda_i(s) ds$ such that 
$M_i(t) = N_i(t) - \Lambda_i(t)$ are martingales. Let further
$N(t)=(N_1(t),...,N_n(t))^T$ be the n-dimensional counting process,
$\Lambda (t)=(\Lambda_1(t),...,\Lambda_n(t))^T$ is its
compensator, such that 
$M(t)=(M_1(t),...,M_n(t))^T$ is an n-dimensional martingale,
and 
define matrices 
$X(t)=(X_{1}(t),\ldots , X_{n}(t))^{T}$ and  
$Z(t)= ( Z_1(t),\ldots, Z_n(t))^T,$ with dimensions $n\times p$ and $n\times q$, respectively.
The individual entry times are summarised in one vector
$a_\bullet=(a_1,\dots, a_n)$.
A superscript $a>0$ denotes a shift in the argument, i.e,
for a generic function $f$, $f^{a}(y)=f(y+a)$.
 For a generic matrix  $C(t)$, with $n$ rows $C_i(t)$, and a n-dimensional vector $v$, $C^v(t)$ is defined through shifting the rows: $C_i^{v}(t)=C_i(t+v_i)$.
 For a generic   matrix $C$, a minus superscript, $C^-$, denotes the  Moore-Penrose  inverse. 
An integral, $\int$, with no limits denotes integration over the whole range.

\section{Identification of the entering nonparametric parameters}\label{sec:identification}
In many cases some covariates will enter both the $X $ and the $Z$ design. If this is the case, then the functions $\alpha$ and $\beta$ 
are not  identified in model \eqref{model} -- constants can be shifted for the components that share the same covariate without altering the intensity.
Without loss of generality we assume that  $X$ and $Z$ share the first $d$  $(0\leq d \leq \min(p,q))$  columns, i.e., for all $i=1,\dots, n$, 
\[
X_{il}=Z_{il}, \quad l\leq d.
\]
We formulate the problem using group-theoretic arguments, see also \cite{Carstensen:07, Kuang:etal:08}. Fix constants
$c_1,\dots, c_d$ and define
$f_l$ as  $\Re^{p+q}$ valued function having all entries but the $l'th$ and the $(d+l)'th$ equal zero:
\[
f_l(s,u)=\left(0, \cdots ,0, c_ls, 0, \cdots, 0,-c_l(u-a_0), 0,\cdots,0)\right)^T, \ (l=1,\dots,d).
\]
We  define the group $G$ by
\begin{align}\notag
G=\left\{g: \begin{pmatrix} A\\ B\end{pmatrix} \mapsto 
\begin{pmatrix} 
A\\ 
B
\end{pmatrix}
+ h\  | \quad h \in Lin(f_1, \dots f_d)\right\}.
\end{align}
The identification problem can be rephrased as that
the intensity defined in \eqref{model} is a function of $(A,B)^T$, which is invariant to transformations $g \in G$.
In the sequel we circumvent the identification issue by adding the following constraint
\begin{align}\label{eq:identification}
A_l(t_{max})=\int_{0}^{t_{max}} \alpha_l(s) \mathrm ds =0, \quad  (l=1,\dots,d),\end{align}
noting that for any solution $(A_0,B_0)$ of model \eqref{model}, there exists a unique solution $(A,B)=g(A_0,B_0)$ that fulfills
\eqref{eq:identification}.
Clearly other choices are also possible.

\section{Least squares minimisation ignoring the identificating of the nonparametric parameters}

We split the identification challenge in two. First we estimate ignoring identification of the parameters, and then we show in next section how to identify the estimated parameters.
In this section we therefore ignore the identification problem keeping in mind that the solutions below are hence not unique. 
We motivate our estimator $(\widehat A, \widehat B)$ 
via the following least squares criteria.
\begin{align*}
\arg \min_{\overline A, \overline B
}\sum_i \int \left\{ \int_0^t \mathrm dN_i(s) - \sum_j \int_0^t X_{ij}(s) d \overline  A_j(s) - \sum_k \int_0^t Z_{ik}(s) d \overline  B_k^{a_i}(s) 
\right\}^2  \mathrm dt,
\end{align*}
where the integrals can be understood as Stieltjes integrals, noting that $X_i$ and $Z_i$ are left continuous.
Minimisation runs over all possible integrators.
One can already see that the minimiser, if it exists, will be a  step-function,  since $\int_0^t \mathrm dN_i(s) $ is a step function.
To simplify notation we will generally work in matrix notation so that above minimisation criteria can also be written as
\begin{align*}
\arg \min_{\overline A, \overline B
}\sum_i \int \left\{ \int_0^t \mathrm dN_i(s) - \int_0^t X_i(s) d \overline  A(s) - \int_0^t Z_i(s) d \overline  B^{a_i}(s) 
\right\}^2  \mathrm dt.
\end{align*}

Straight forward computations utilzing calculus of variations lead to $(\widehat A, \widehat B
)$ solving the following  first order conditions for all $t\in [0,t_{max}]$, $a \in [a_0,a_{max}]$:
\begin{align*}
 & \sum_i X_i(t)^T \left \{ dN_i(t)  -  X_i(t) d\widehat A(t) -   Z_i(t) 	\mathrm d \widehat B^{a_i}(t) 
  \mathrm dt \right \}=0,\\
& \sum_i  Z_i^{-a_i}(a)^T \left\{ dN_i^{-a_i}(a)  -  Z_i^{-a_i}(a) d \widehat B(a) - X_i^{-a_i}(a) 	\mathrm d \widehat A^{-a_i}(a)
\ \right\} =0.
\end{align*}
Rearranging yields
\begin{align*}
&  \sum_i X_i(t)^T  dN_i(t)   -     \sum_i X_i(t)^T  Z_i(t) 	\mathrm d\widehat B^{a_i}(t) 
=  X(t)^T   X(t) \mathrm d\widehat  A(t),\\
 &\sum_i  Z_i^{-a_i}(a)^T dN_i^{-a_i}(a)  -   \sum_i  Z_i^{-a_i}(a)^T X_i^{-a_i}(a) 	\mathrm d\widehat  A^{-a_i}(a) 
 =   Z^{-a_\bullet}(a)^T Z^{-a_\bullet}(a) d \widehat  B(a).
\end{align*}
The last set of equations can be further  rewritten to the backfitting equations
\begin{align} \label{bf1}
 \widehat  A(t) 
            &=  \int_{0}^t X(s)^{-} dN(s) - \int  E_1(t|u)  d\widehat  B(u)
\\
\widehat B(a) 
                & = \int_{a_0}^a Z^{-a_\bullet}(u)^{-} dN^{-a_\bullet}(u) - \int E_2(a|s)  d \widehat  A(s),
                \label{bf2}
\end{align}
where
\begin{align*}
  E_1(s|u) & =  \sum_i  \{X^{T}(u-a_i)X(u-a_i)\}^{-1}X_i^{-a_i, T}(u)  Z^{-a_i}_i(u)   I(a_i     \leq u \leq a_i+s), \\
  E_2(u|s) & =  \sum_i  \{Z^{-a_\bullet,T}(s+a_i)Z^{-a_\bullet}(s+a_i)\}^{-1}Z_i^T(s)X_i(s) I(a_0-a_i \leq s \leq u-a_i).  
 \end{align*}

\begin{remark}\label{remark:simpleE}
In the case with no covariates, i.e., 
\[
\lambda_i(t) =  Y_i(t)\{ \alpha (t) +  \beta(a_i+t)\},
\]
with $X_i(s)=Z_i(s)=Y_i(s) \in \Re$, 
the risk indicators are 
\begin{align*}
  E_1(s|u)  & =  \sum_i  \frac{1}{\sum_{i'} Y_{i'}(u-a_i)}   Y_i^{-a_i}(u)   I(a_i     \leq u \leq a_i+s), \\
  E_2(u|s)  & =  \sum_i  \frac{1}{\sum_{i'} Y^{-a_{i'}}_{i'}(s+a_i)} Y_i(s) I(a_0-a_i \leq s \leq u-a_i).
\end{align*}
\end{remark}

\section{Establishing existence, identification and uniqueness  of the estimator}
In section \ref{sec:identification} we outlined the identification problem but ignored it when establishing the estimator in the previous section. In this section we provide a fully identified estimator of our problem.
When aiming to solve  equations \eqref{bf1} and \eqref{bf2} the identification problem can no longer be ignored.
In order to get a better grip of the situation we will now rewrite the backfitting equations as a linear operator equation.
We can compress equations \eqref{bf1} and \eqref{bf2} into one matrix equation:
\[
 \begin{pmatrix}
            \widehat A \\
           \widehat B\\
         \end{pmatrix}= 
\begin{pmatrix}
  \int_{0}^t X(s)^{-} dN(s) \\
\int_{a_0}^a Z^{-a_\bullet}(u)^{-} dN^{-a_\bullet}(u) 
         \end{pmatrix} +
\begin{pmatrix}
   0  &- E_1 \\
              - E_2&  0
         \end{pmatrix}\times  \begin{pmatrix}
        \widehat    A \\
       \widehat    B\\
         \end{pmatrix},
\]
where with some miss-use of notation $E_l f(\cdot)=\int E_l (\cdot,y)f(y) \mathrm dx, (l=1,2)$. Or even simpler
\begin{align}\label{eq:operator1}
\widehat \theta= \widehat m + E\widehat \theta,
\end{align}
with obvious notation, and linear operator $E$:
\[
\widehat \theta=   \begin{pmatrix}
        \widehat    A \\
       \widehat    B\\
         \end{pmatrix}, \quad  \widehat m=  \begin{pmatrix}
     \int_{0}^t X(s)^{-} dN(s) \\
\int_{a_0}^a Z^{-a_\bullet}(u)^{-} dN^{-a_\bullet}(u) 
         \end{pmatrix}, \quad  E= \begin{pmatrix}
   0  &- E_1 \\
              - E_2&  0
         \end{pmatrix}.
\]
Note that $\widehat m$ is composed of the marginal Aalen estimators of the two time scales,  $t$ and $a$.
Additionally, the operator $E$ is compact because it is the composition of an integral operator, which is compact, and a derivative operator, which is bounded.
The operator $E$ being compact means that it can be arbitrarily close approximated by a  finite dimensional matrix which simplifies both the numerical and theoretical considerations.
If the eigenvalues of $E$ are bounded away from one, then,  $(I-E)$ is invertible and we have
\[
\widehat \theta= (I-E)^{-1} \widehat m.
\]
Hence existence and uniqueness of our proposed estimator can be translated to properties of the eigenvalues of $E$.
One can for instance easily verify that if some covariates are both in the $X $ and the $Z$ design, then $E$ will have an eigenvalue equal to one -  as discussed in the following remark.
\begin{remark}
Consider the most simple case $1=d=p=q$, i.e.,  $\lambda_i(t) =  Y_i(t)\{ \alpha (t) +  \beta(a_i+t)\}$.
Given a constant $c\in \Re $,
consider the  pair of linear function $f_1=(f_{11},f_{12})^T$ with $f_{11}(s)=cs, \ f_{12}(u)=-c(u-a_0)$, as defined in Section \ref{sec:identification}.
Assuming that $\sum Y_i(s)$ and $\sum Y_i(u-a_i)$ are bounded away from zero on the whole range
$s\in [0, t_{max}], u\in [a_0, a_{max}]$,
one can easily verify that
\begin{align*}
E_2f_{11}(u)&=c \int E_2(u|s) \mathrm ds= c(u-a_0),\\
E_1f_{12}(s)&=-c \int E_1(s|u) \mathrm du= -cs.
\end{align*}
To see this, e.g., for the second equation,  note
\[
\int E_1(s|u) \mathrm du=  \sum_i   \int_{a_i}^{a_i+s}\frac{1}{\sum_{i'}Y_{i'}(u-a_i)}   Y_i^{-a_i}(u)  \mathrm du
=\int_0^s \frac{\sum_i Y_i(t) }{\sum_{i'}Y_{i'}(t)} \mathrm dt=s.
\]
Hence, we have 
\[
E \begin{pmatrix}f_{11}\\ f_{12}\end{pmatrix}=\begin{pmatrix}-E_1f_{12}\\- E_2f_{12}\end{pmatrix}=
\begin{pmatrix}f_{11}\\ f_{12}\end{pmatrix}.
\]
So that one is clearly an eigenvalue of $E$ with corresponding eigenfunction $f_1=(f_{11},f_{12})^T$.
In other words the identification issue of the model carries over to the estimator.
With analogue arguments one can show that in the more general case the eigenspace corresponding
to eigenvalue equal one includes the functions in $Lin(f_1, \dots f_d)$.  Functions  $f_1,\dots, f_d$ are  defined in Section \ref{sec:identification}.
\end{remark}
We now utilize constraint \eqref{eq:identification}  and incorporate it into  new backfitting equations:
\begin{align} \label{bf11}
 \widehat  A(t) 
            &=  \int_{0}^t X(s)^{-} dN(s) - \int  E_1(t|u)  d\widehat  B(u),
\\
\widehat B(a) 
                & = \int_{a_0}^a Z^{-a_\bullet}(u)^{-} dN^{-a_\bullet}(u) - \int E_2(a|s)  d \widehat  A(s) + \frac{\widehat A^{d_q}(t_{max})}{t_{max}} (a-a_0), \label{bf22}
   \end{align}
   where $\widehat A^{d_q}$ is the q-dimensional vector  $\widehat A^{d_q}= (A_1, \dots, A_d, 0, \dots, 0)^T$.
  This translates to the new operator equation
  
   \begin{align}\label{eq:operator2}
\widehat \theta= \widehat m + \overline E\widehat \theta, \quad  \overline E= \begin{pmatrix}
   0  &- E_1 \\
              - \overline E_2&  0
         \end{pmatrix},
\end{align}
where $\overline E_2 h (a)= \int E_2 (a|s)dh(s)- (a-a_0) h^{d_q}(t_{max}) t_{max}^{-1}$.
The next proposition states that the solutions of $\eqref{eq:operator2}$ 
include all relevant solutions of \eqref{eq:operator1}
and that every solution of $\eqref{eq:operator2}$ is a solution of $\eqref{eq:operator1}$.
\begin{prop}\label{prop:operator1}
For every solution $\widehat \theta$ of  \eqref{eq:operator1}, define
\[
\widehat {\theta}_0=(I-\widetilde \Pi )\widehat \theta,
\]
where 
\[ 
\widetilde \Pi \begin{pmatrix}
h_1 (t)\\ h_2(a)
\end{pmatrix}
= 
\begin{pmatrix}
t h_1^{d_p}(t_{max}) t_{max}^{-1} \\ -(a-a_0) h_1^{d_q}(t_{max}) t_{max}^{-1}
\end{pmatrix}.
\]
Then
$\widehat {\theta}_0$ is a solution of $\eqref{eq:operator2}$ and
\begin{align}\label{directsum}
\widehat \theta_0 +  Lin(f_1, \dots f_d), 
\end{align}
are further solutions of \eqref{eq:operator1}. Reversly, for every solution ${\widehat \theta}_0$
of \eqref{eq:operator2},
all functions of the form \eqref{directsum} are solutions of  \eqref{eq:operator1}.
\end{prop}
The proof can be found in the appendix.

With Proposition \ref{prop:operator1} at hand it is justified to define our estimator as the solution of
\eqref{eq:operator2}.
We will now discuss  existence and uniqueness of the solution of \eqref{eq:operator2}.

Note that $E$ is known and hence one can calculate a numerical approximation of its eigenvalues by working on a grid.
Consider the sub-space
\begin{align*}
K=\{h=(h_1,\dots, h_d,0,\dots,0) | \ h_l: \Re  \to \Re , \ x \mapsto c_lx , \  c_l\in \Re , \ l=1,\dots, d\}.
\end{align*}
It holds that  $\overline E_2=E_2(I-\Pi)$, where $\Pi$ is a projection into $K$.
We have $K \subseteq kern(I-E_2)$.
We can  check whether $K$ equals $kern(I-E_2)$.
This can be done by calculating the dimension of the eigenspace of $E_2$ corresponding to an eigenvalue equal one.
The dimension will be at least $d$. If it is exactly $d$, then $K=kern(I-E_2)$.

The  next proposition states that if $kern(I-E_2)=K$,
and $kern(I-E)=Lin( f_1,\dots,f_d)$,
 then both  $I-\overline E_2$
and  $I-\overline E$ are bijective.

\begin{prop}\label{prop:operator2}
Assume that $E_2$ has Eigenvalue 1 with multiplicity $d$.
Then, $(I-\overline E_2)$ will be bijective.
If furthermore E has Eigenvalue 1  with multiplicity $d$,
then $(I-\overline E)$ is bijective and hence invertible. In particular a solution of equations \eqref{eq:operator2} exists and it is unique.
\end{prop}
The proof can be found in the Appendix.

 \section{Calculating the estimator}
 There are two major ways of calculating the proposed estimator.
 Either one directly calculates $(I-\overline E)^{-1}$ and applies it on $\widehat \theta$ or something closer to an iterative procedure.
 For the latter,
 by iterative application of \eqref{eq:operator2} we derive that
\begin{align}\label{infinitesum}
 \widehat \theta= \sum_{r=0}^\infty \overline E^r (\widehat m ) +\overline E^\infty(\widehat \theta).
\end{align}
If the absolute values of the  eigenvalues of $\overline E$
are bounded from above by a constant strictly smaller than 1, then \eqref{infinitesum} is well defined with $E^\infty=0$, and the converging series
 \[
 \widehat \theta= \sum_{r=0}^\infty \overline E^r (\widehat m ),
 \]
 so that the iterative algorithm
 \begin{align}\label{bf}
\widehat \theta^{(r)}= \widehat m + \overline E\widehat \theta^{(r-1)} 
\end{align}
converges from any starting point. 
Note that \eqref{bf} is the usual way the backfiting equations \eqref{bf11},\eqref{bf22} or equivalently \eqref{eq:operator2}  are solved.
Another way is to  calculate the finite sum 
\[
\widetilde \theta= \sum_{r=0}^{\overline r} \overline E^r (\widehat m ),
\]
with some stopping criteria $\overline r$.
We conclude 
that the proposed estimator can be calculated in a straight forward manner from
the compound Aalen estimator $\widehat m$ and the operator $\overline E.$

We now briefly  discuss how $\overline E$ can be calculated in the simple case $1=d=p=q$.
Here, $\overline E$ can be approximated by a $j\times k$ matrix where $j, k$ are the number of grid points 
in $[0,t_{max}]$ and $[a_0, a_{max}]$, respectively.
This is done by first calculating the values $E_1(s_0,a_0)$ and $E_2(a_0,s_0)$ for every grid point; see Remark \ref{remark:simpleE} for the definitions of the the functions.
We call the resulting matrices $E_1^{mx}$ and $E_2^{mx}$.
Afterwards, $\overline {E}_2^{mx}$ is derived from $E_2^{mx}$, via
\[
  \overline E_2^{mx}= E_2^{mx}+  \begin{pmatrix}
   0  & \cdots & 0& s_1/s_j \\
           0&\dots&0&  s_2/s_j\\
            \vdots &&\vdots&\vdots \\
                       0&\dots&0&  1
         \end{pmatrix}.
\]
The matrices are then transformed to the desired operator via

\begin{align*}
\Delta=\begin{pmatrix}
1  & -1     & 0 &  \cdots     & 0      \\
0  & \ddots & \ddots &    \ddots    & \vdots \\
\vdots  & \ddots & \ddots & \ddots       & 0 \\
\vdots & \ddots & \ddots & \ddots &-1    \\
0   & \cdots & \cdots & 0&1  \\
\end{pmatrix},\quad 
E_1^{op}=E_1^{mx} \times \Delta, \quad 
\overline E_2^{op}= \overline E_2^{mx}\times \Delta.
       \end{align*}
Finally,
\begin{align*}
\overline  E^{op}= \begin{pmatrix}
   0  &- E_1^{op} \\
              - \overline E_2^{op}&  0
         \end{pmatrix}.
       \end{align*}
 So that given a function  $h: [0,t_{max}]\times [a_0, a_{max}] \rightarrow \Re $,
one calculates its values on the grid and summarises it in a vector
  $ h^{grid}$. The function 
$ \overline Eh $ is then approximated via $\overline E^{op}h^{grid}$ where the latter is a simple matrix multiplication.
\section{Asymptotics}
Note that we  have
\begin{align}\label{true:eq}
 \theta=  m + \overline E\theta,
\end{align}
where $m$ arises from $\widehat m$ by replacing $N$ by $\Lambda$. It is hereby quite remarkable that $\overline E$
is the observable operator from the previous sections and not some asymptotic limit.
We further conclude  that the least square solution \eqref{bf11} and \eqref{bf22} is  a plug-in estimator of \eqref{true:eq}.
The estimation error is then given as
\begin{align}\label{error:eq}
\widehat \theta - \theta= \widehat m -m + \overline E(\widehat \theta- \theta).
\end{align}
As in the last section, 
If $\overline E$ has eigenvalues all bounded away from  one, then 
\[
\widehat \theta - \theta= (I-\overline E)^{-1} (\widehat m -m).
\]
So the asymptotic behaviour of $\widehat \theta - \theta$ can be deduced from the asymptotic behaviour
of $(I-\overline E)^{-1}$ and $(\widehat m -m)$, with the latter being the compound estimation error of  two additive Aalen models on different time-scales.

\begin{theorem}\label{thm:asymptotics}
Under assumptions (A)--(G),  the estimator $\widehat \theta$ exists.  
Furthermore  the estimator $\widehat \theta$ is $n^{1/2}$ consistent:
\[
n^{-1/2}  (\widehat \theta - \theta )\rightarrow (I-\widetilde E)^{-1}U,
\]
in Skorohod space $D^{p+q}[0,a_{max}]$. 
Here,  $(\widehat \theta - \theta )$ is treated as one stochastic process defined on $[0,a_{max}]$
by setting for $j=1,\dots, p$ and $\nu \in [t_{max}, a_{max}]$, $(\widehat \theta - \theta )_j (\nu) =(\widehat \theta - \theta )_j(t_{max})$. And similarly, for $j=p+1, \dots, p+q$ and $\nu \in [0, a_{0}]$, $(\widehat \theta - \theta )_j (\nu) =0$.
The process $U$ is a $p+q$ dimensional mean-zero Gaussian process with covariation matrix $\Sigma(\nu_1,\nu_2)$ described in the Appendix, 
and $\widetilde E$ is the  limit of $\overline E$.

\end{theorem}
The proof can be found in the Appendix.

\section{Confidence Bands}
While we could use the central limit theorem of the previous section to construct confidence bands, 
it has been suggested that better small sample performance can be achieved by directly
bootstrapping the estimation error.
We propose a wild bootstrap approach based on the relationship
\begin{align*}
\widehat \theta - \theta=(I-\overline E)^{-1} (\widehat m -m)&=(I-\overline E)^{-1}  \begin{pmatrix}  \int_{0}^t X(s)^{-} dM(s)   \\
\int_{a_0}^a Z^{-a_\bullet}(u)^{-} dM^{-a_\bullet}(u)
\end{pmatrix}
\\ &= (I-\overline E)^{-1} \begin{pmatrix} \mathcal M_1  \\
\mathcal M_2
\end{pmatrix} 
\end{align*}
Since ($I-\overline E)^{-1} $ is known,
it is enough to to only approximate $\mathcal M$.
We do this via the wild bootstrap version
\[
\widehat {\mathcal M}^{(1)}= \begin{pmatrix}  \int_{0}^t X(s)^{-} d\widetilde {N}(s)   \\
\int_{a_0}^a Z^{-a_\bullet}(u)^{-} d\widetilde {N}^{-a_\bullet}(u)
\end{pmatrix}, \quad \widetilde N_i(s)= G_i N_i(s),
\]
or
\begin{align*}
\widehat {\mathcal M}^{(2)}&= \begin{pmatrix}  \int_{0}^t X(s)^{-} d\widetilde {M}(s)   \\
\int_{a_0}^a Z^{-a_\bullet}(u)^{-} d\widetilde {M}^{-a_\bullet}(u)
\end{pmatrix}, \\ \quad \int_0^t \widetilde M_i(s)\mathrm d s&= G_i \left( \int_0^t N_i(s) \mathrm ds -\big(\int_0^t (X_i(s)\mathrm d\widehat A(s)+ \int_0^t Z_i(s) \mathrm d \widehat B(s+a_i)\big)\right),
\end{align*}
where $G_i$ is a mean zero random variable with unit variance.
The random variable $G_i$ is generated such that for fixed $i$, it is independent to all other variables.
It is straight forward to confirm that $\widehat {\mathcal M}^{(r)}, \ r=1,2$ is a mean zero process that has the same covariance  as
$\mathcal M$ (The covariance of $\mathcal M$ is given in the appendix).
Hence, we directly derive the following proposition.
\begin{prop}\label{prop:bootstrap}
Under assumptions (A)--(G), the bootstrapped estimation error is uniformly consistent, i.e.,  for $r=1,2$
\[
n^{-1/2}  ((I-\overline E)^{-1}\widehat {\mathcal M}^{(r)} )\rightarrow (I-\widetilde E)^{-1}U,
\]
in Skorohod space $D^{p+q}[0,T]$,
where $U$ is is described in Theorem 1. 
\end{prop}
The proof can be found in the Appendix.

One useful consequence of this is that we can estimate standard errors of our estimator $\hat \theta$ based on
the approximation from the bootstrap. We denote these estimators as $\hat \sigma_r(t)$ for the two components $r=1,2$.

\begin{corollary}\label{cor:bootstrap}
Under assumptions (A)--(G), the bootstrapped errors lead 
to confidence bands $CB^{(r)}$ for  ${\theta}(\nu)$ over $\nu\in[\nu_1,\nu_2]$ providing an asymptotic coverage probability of $1 - \alpha$, where
\[
	CB^{(r)}(\nu)= \theta(\nu) +/- c_{1-\alpha} \hat \sigma_r(\nu),
\]
and
\[
c_{1-\alpha} = (1-\alpha)\quad  \textrm{quantile of } \quad \mathcal L\left\{ \sup_{[\nu_1,\nu_2]}n^{-1/2}  
	\frac{\abs{(I-\overline E)^{-1}\widehat {\mathcal M}^{(r)} }}{\hat \sigma_r} | X,Z, N\right\}
\]
\end{corollary}

We explore the performance of the estimator of the standard error and the uniform bands in the next section.

\section{Simulations}

We generated data from the simple two-time scale model with age and duration 
that resemble the data we consider in worked example in the next section. 
Thus assuming that the hazard for those under risk is given as 
$\beta(t+a_i)+\alpha(t)$, where $\beta(a) \equiv 0.067$ and the entry ages where 
drawn uniformly from $[0,25]$ but making sure that 10 \% of the data started in $0$ to 
(to avoid difficulties with left truncation in the estimation). 
The $\alpha(t)$ component was piecewise constant
with rate $0.32$ in the time-interval $[0,0.25]$, then $0.48$ in  $(0.25,0.5]$ and then finally to 
satisfy our constraint $-0.044$ in $(0.5,5]$, so that $\int_0^5 \alpha(s) ds =0$. 
All subjects were censored after $5$ years of follow up. 

In all simulations we 
used a discrete approximation based on a time-grid of 
either 100 points in both the age direction $[0,30]$ and on the duration 
time-scale $[0,5]$.

\subsection{Bias of backfitting}

We considered sample sizes 100, 200 and 400 and show the bias for 
the two-components in 
Table 1 based on 1000 realizations. 

\begin{table}[ht]
\begin{tabular}{l| c c c} 
age & n=100&n=200&n=400 \\ \hline
\hline
\hline
 $ 6.717 $&$ -0.001 $&$ 0.006 $&$ -0.004 $ \\
 $ 13.788 $&$ 0.009 $&$ 0.003 $&$ -0.006 $ \\
 $ 20.859 $&$ 0.018 $&$ 0.001 $&$ 0.002 $ \\
 $ 27.929 $&$ 0.027 $&$ 0.004 $&$ 0.010 $ \\
 $ 35 $&$ 0.078 $&$ 0.006 $&$ 0.013 $ \\
\hline
\hline
time & n=100&n=200&n=400 \\ \hline
\hline
 $ 0.96 $&$ 0.018 $&$ 0.009 $&$ 0.006 $ \\
 $ 1.97 $&$ 0.015 $&$ 0.007 $&$ 0.005 $ \\
 $ 2.98 $&$ 0.009 $&$ 0.005 $&$ 0.003 $ \\
 $ 3.99 $&$ 0.005 $&$ 0.002 $&$ 0.002 $ \\
 $ 5 $&$ 0 $&$ 0 $&$ 0 $ \\
\end{tabular}
\caption{Bias of backfitting algorithm for sample sizes $n = 100, 200, 400$ for the 
age and time component for selected ages and time points. 
	Based on 1000 realisations.  
}
\label{tab:tab1}
\end{table}

We note that the the backfitting algorithm is almost unbiased across all sample
size and improves as the sample size increases.  This is despite the fact that
the simulated component in the time-direction really is quite wild. 

\subsection{Bootstrap uncertainty}

Secondly, we demonstrate that our bootstrap seems to work well to describe the uncertainty of the
estimates.  We simulated data as before and based on 1000 realisations with 100 bootstrap's based
on $G_i dN_i$ we estimated: a) the point-wise standard error for the two-components; b) computed the 
pointwise coverage baed on these; c) and constructed
uniform confidence bands, as described in Corollary 1, for the the two components and its coverage. 

\centerline{Table 2 around here} 

\begin{table}
	\begin{tabular}{ l| c c c c|| c c c c } 
n &age&mean se&sd&cov&time&mean se&sd&cov \\\hline
\hline
\hline
 $ 100 $&$ 6.717 $&$ 0.224 $&$ 0.231 $&$ 0.912 $&$ 0.96 $&$ 0.044 $&$ 0.045 $&$ 0.954 $ \\
 $ 100 $&$ 13.788 $&$ 0.297 $&$ 0.298 $&$ 0.935 $&$ 1.97 $&$ 0.039 $&$ 0.04 $&$ 0.946 $ \\
 $ 100 $&$ 20.859 $&$ 0.351 $&$ 0.357 $&$ 0.943 $&$ 2.98 $&$ 0.032 $&$ 0.034 $&$ 0.951 $ \\
 $ 100 $&$ 27.929 $&$ 0.391 $&$ 0.402 $&$ 0.938 $&$ 3.99 $&$ 0.024 $&$ 0.024 $&$ 0.966 $ \\
 $ 100 $&$ 35 $&$ 0.460 $&$ 0.464 $&$ 0.932 $&$ 5 $&$ 0.016 $&$ 0.017 $&$ 0.874 $ \\
\hline
\hline
 $ 200 $&$ 6.717 $&$ 0.158 $&$ 0.155 $&$ 0.94 $&$ 0.96 $&$ 0.031 $&$ 0.031 $&$ 0.951 $ \\
 $ 200 $&$ 13.788 $&$ 0.207 $&$ 0.206 $&$ 0.942 $&$ 1.97 $&$ 0.027 $&$ 0.027 $&$ 0.960 $ \\
 $ 200 $&$ 20.859 $&$ 0.243 $&$ 0.237 $&$ 0.948 $&$ 2.98 $&$ 0.022 $&$ 0.022 $&$ 0.966 $ \\
 $ 200 $&$ 27.929 $&$ 0.271 $&$ 0.262 $&$ 0.945 $&$ 3.99 $&$ 0.017 $&$ 0.017 $&$ 0.972 $ \\
 $ 200 $&$ 35 $&$ 0.328 $&$ 0.329 $&$ 0.933 $&$ 5 $&$ 0.011 $&$ 0.012 $&$ 0.933 $ \\
\hline
\hline
 $ 400 $&$ 6.717 $&$ 0.114 $&$ 0.118 $&$ 0.948 $&$ 0.96 $&$ 0.022 $&$ 0.022 $&$ 0.951 $ \\
 $ 400 $&$ 13.788 $&$ 0.148 $&$ 0.153 $&$ 0.946 $&$ 1.97 $&$ 0.019 $&$ 0.019 $&$ 0.957 $ \\
 $ 400 $&$ 20.859 $&$ 0.173 $&$ 0.18 $&$ 0.937 $&$ 2.98 $&$ 0.015 $&$ 0.015 $&$ 0.960 $ \\
 $ 400 $&$ 27.929 $&$ 0.192 $&$ 0.196 $&$ 0.943 $&$ 3.99 $&$ 0.012 $&$ 0.012 $&$ 0.970 $ \\
 $ 400 $&$ 35 $&$ 0.235 $&$ 0.245 $&$ 0.934 $&$ 5 $&$ 0.008 $&$ 0.008 $&$ 0.950 $ \\
\end{tabular}
\caption{
	Uncertainty estimated from bootstrap for sample sizes $n = 100, 200, 400$ for the age and time component for selected ages and time points. 
	Based on 1000 realisations and a bootstrap with 100
repetitions. mean of estimated standard errors (mean se), standard deviation of estimates (sd) and 95 \% pointwise coverage (cov). 
}
\label{tab:tab2}
\end{table}

We note that the standard error is well estimated by the bootstrapped standard deviation across all
sample sizes and for both components. In addition the pointwise coverage is 
close to the nominal 95 \% level for the larger sample sizes. But even for $n=100$ the coverage is 
reasonable for most time-points for the two components. 

Finally, we also considered the performance of the confidence bands based on our bootstrap approach.

\centerline{Table 3 around here} 

\begin{table}
\begin{tabular}{ l| c c } 
n	&coverage (age) & coverage (time) \\\hline
\hline
 $ 100 $&$ 0.797 $&$ 0.792 $ \\
 $ 200 $&$ 0.912 $&$ 0.915 $ \\
 $ 400 $&$ 0.952 $&$ 0.939 $ \\
\hline
\end{tabular}
\caption{Coverage of confidence bands estimated from bootstrap for sample sizes $n = 100, 200, 400$ for the age and time component. 
Based on 1000 realisations and a boostrap with 100 repetitions. 
}
\label{tab:tab3}
\end{table}

When $n$ gets larger these bands are quite close to the nominal 95 \% level, 
but for $n=100$ the asymptotics have not quite set in to make the 
entire band work well.

\section{Application to the TRACE study}

The TRACE study group (see e.g. \cite{trace} ) has
collected information on more than 4000 consecutive patients with
acute myocardial infarction (AMI) with the aim of 
studying the prognostic importance of various risk 
factors on mortality. We here consider a subset of 1878 of these
patients that are available in the timereg R package. 
At the age of entry (age of diagnosis) the
patients had various risk factors recorded, but we here just show the
simple model with the effects of the two-time-scales age and duration. 
It is expected that the duration time-scale has a strong initial effect of dying that then
disappears when patients survive the first period right after their AMI.



We then estimated the two-time-scale model $\alpha(t)+\beta(t+a_i)$ under the 
identifiability condition 
that $\int_0^5 \alpha(s) ds=0$. Restricting attention to patients more than 
40 years of age, and within the first 5 duration years after the diagnosis. 

First we estimate the mortality on the two time-scales separately, the two
marginal estimates, see Figure 1. Panel (a) shows the cumulative hazard on the
age time-scale with the marginal estimate (full line) and the one with
adjustment for duration effects (broken line), and panel (b) the mortality on
the duration time-scale with the marginal estimate (full line) and with
adjustment for age effects (broken line).  We note that on the duration
time-scale the cumulative hazard is quite steep.  In addition we show 95 \%
confidence bands based on our bootstrap (regions), and the pointwise confidence
intervals (dotted line). 

\centerline{Figure 1 about here}

\begin{figure}[ht]
  \centering
  \includegraphics{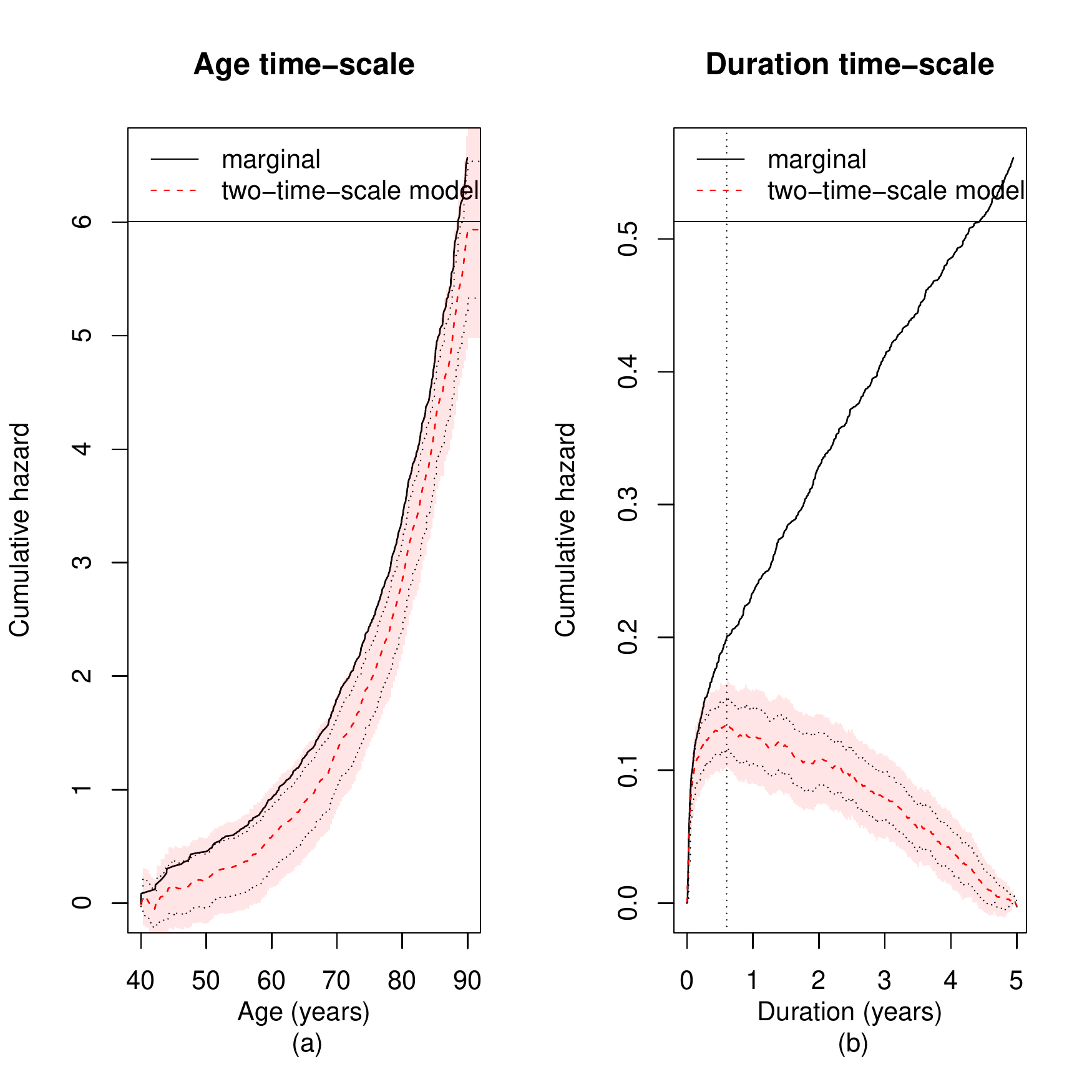}
      \caption{Cumulative baseline on the two time-scales estimated marginally (full line) and in the two-time-scale model (broken line). 
	 Confidence bands (regions) and pointwise confidence intervals (dotted lines).}
  \label{figt:tracefig}
\end{figure}

Taking out the duration effect slightly alters the estimate of the age-effect.  In contrast the 
duration effect is strongly confounded by age effect estimates, and here the two-time scale model more
clearly demonstrates what is going on on  the duration time-scale. The duration effect is strong initially and then
after surviving the first 220 days we see a protective effect (dotted vertical line).

We stress that the interpretation of the hazards on the two-time scales are difficult, due to, for example, the
constraint that needs to be imposed to identify a specific solution. Nevertheless, it very useful to see the components 
from the two time-scales that jointly make up the hazard for an individual, and can be used for the prediction purposes 
as we demonstrate further below.  Note also that due to the additive structure the duration effect can be 
interpreted as giving relative survival due to the duration time-scale. 

\centerline{Figure 2 about here}

\begin{figure}[ht]
  \centering
  \includegraphics{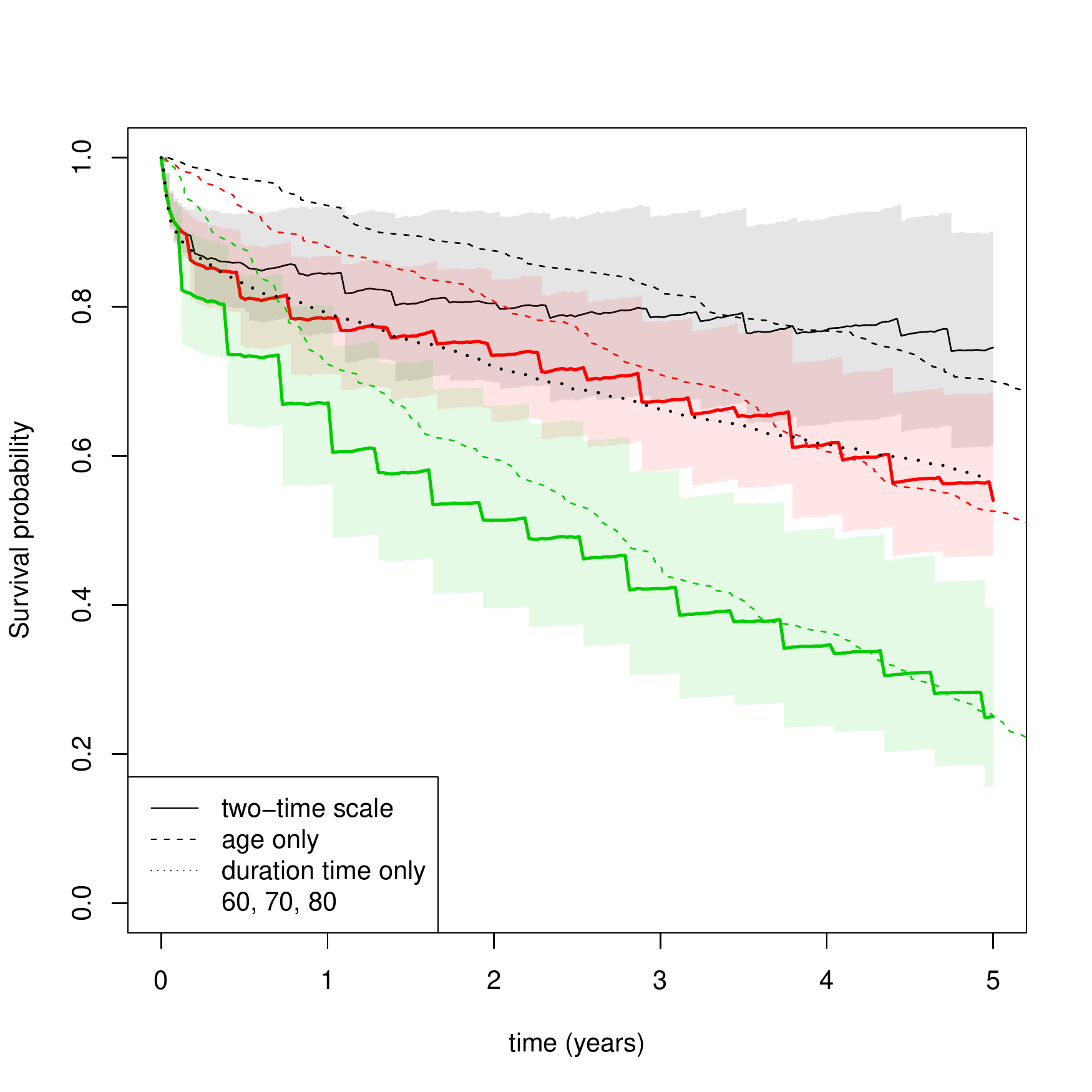}
	\caption{Predicted survival with 95 \% confidence bands (regions) for a subject that is 60,70, and 80, respectively (full lines). 
	Predicted survival using only age for the three ages (broken lines), and survival using only duration (dotted line). }
  \label{figt:survtracefig}
\end{figure}

In Figure 2 we show the survival predictions for subjects that are 60, 70, or
80, respectively, using the two-time scale model.  Thus computing 
$\exp(- (\hat B(a_0+t) - \hat B(a_0)) + \hat A(t))$ and constructing the 
confidence bands using the bootstrap approach for 
$(\hat B(a_0+t) - \hat B(a_0)) + \hat A(t)$  for $t \in [0,5]$.
These curves are a direct consequence of having the two-components and
are directly interpretable.

\section{Discussion}

By utilising the additive structure we have demonstrated that one can estimate
the effect of two time-scales directly by a backfitting algorithm that does not
involve smoothing. By working on the cumulative this also lead to uniform
asymptotic description and a simple bootstrap procedure for getting estimates of the 
uncertainty and for constructing for example confidence intervals. 
These cumulative may form the basis for smoothing based estimates when the hazard are
of interest, but often the cumulative are the quantities of key interest for 
example when interest is on survival predictions. 

Clearly, the model could also be fitted by a more standard backfitting approach
working on the hazard scale as in ... for multiplicative hazard models. 

Our backfitting approach can be extended for example 
the age-period-cohort model but here identifiability conditions are more complex
to build into the estimation.

\newpage 
\appendix
\section{Proofs}
\subsection{Proof of Proposition \ref{prop:operator1}}
With 
$f_k(s,u)=\left(0, \cdots ,0, c_ks, 0, \cdots, 0,-c_k(u-a_0), 0,\cdots,0)\right)^T, \ (k=1,\dots,d)$,
the proposition 
directly follows from  $Lin(f_1, \dots f_d)\subseteq Kern(I-E)$,
and the fact that $\widetilde \Pi$ is a projection into $Lin(f_1, \dots f_d)$.

\subsection{Proof of Proposition \ref{prop:operator2}}
Since the eigenspace of $E_2$ corresponding to the eigenvalue equal 1 has dimension $d$, we know its exact form:
\[
\{h=(h_1,\dots, h_d,0,\dots,0) | \ h_l \ \text{is linear}, \  l=1,\dots, d\};
\] 
see also Remark 2.
One can then verify that  $\textrm{kern}(I-E_2)=\textrm{kern}(I-E_2)^l, l=2,3,\dots$.
This is because  linear functions  cannot be constructed as sum of a linear and non-linear functions.
Noting that $E_2$ is a compact operator, we conclude that $I-E_2$ is an isomorphism 
from $\textrm{Im}(I-E_2)$ to $\textrm{Im}(I-E_2)$.
We introduce  the operator $\overline E_2=E_2(I-\Pi)$, where $\Pi$ is the projection onto $\textrm{ker}(I-E_2)$.
The condition $\overline E(h_1,h_2)=(h_1,h_2)$ is equivalent to $\overline E_2E_1 h_2=h_2 $ and  $ E_1\overline E_2 h_1=h_1 $.
Since the eigenspace of $E$ corresponding to an eigenvalue  of 1 has dimension $d$, 
$\overline E(h_1,h_2)=(h_1,h_2)$ is not true for non-linear $h_1, h_2$.
This is because $\overline E=E$ when restricted on non-linear functions $h_1, h_2$.
When considering a linear $h_1$, then $\overline E_2 h_1=0$. We conclude that the solution of
$\overline E(h_1,h_2)=(h_1,h_2)$  is trivial.
Hence  the kern of $(I-\overline E)$ is  trivial. Since $\overline E$ is compact this means $(I-\overline E)$ is bijective, in particular invertible.

\subsection{Assumptions}
We first define a few quantities.\\
For every $\nu$ in, $[0, a_{max}]$, we define the following matrices
\[
R(\nu)=\begin{pmatrix}
   X(\nu) & 0 \\
              0&  Z^{-a_\bullet}(\nu)
         \end{pmatrix}, \quad V(\nu)=(X(\nu),Z(\nu)),
\]
as well as
\begin{align*}
R^{(1)}_{j}(\nu)&=\sum_i R_{ij}(\nu)\\
R^{(2)}_{jk}(\nu)&=\sum_i R_{ij}(\nu)R_{ik}(\nu), \\
V^{(2)}_{jk}(\nu)&=\sum_i V_{ij}(\nu)V_{ik}(\nu), \\
V^{(3)}_{jkl}(\nu)&=\sum_iV_{ij}(\nu)V_{ik}(\nu)V_{il}(\nu).
\end{align*}

We further define
\begin{align*}
\{\widetilde E_1 (s|u)\}_{jk}&=\left( \int h(x) \sum_l \{r^{(2)}(u-x)\}^{-1,T} q^{(2)}(u-x)) I(x\leq u\leq x+s)\mathrm dx\right)_{j, p+k}, \\  \text{for} &\quad j=1,\dots, p, \ k=1,\dots,q\\
\{\widetilde E_2 (u|s)\}_{jk}&= \left(\int h(x) \sum_l \{r^{(2)}(s+x)\}^{-1,T} q^{(2)}(s) I(a_0-x\leq s\leq u-x) \mathrm dx\right)_{p+j, k},
\\
&\quad  \text{for} \ j=1,\dots, \ q, k=1,\dots,p.
\end{align*}
The limiting operator $\widetilde E$ is then defined analogue to $\overline E$ by replacing
$E_1(s|u), E_2(u|s)$ by $\widetilde E_1 (s|u), \widetilde E_2 (u|s)$.
\\
We  make the following assumptions.
\begin{enumerate}
\item[(A)]
There exist continuous functions $r^{(1)}_j,r^{(2)}_{jk},v^{(2)}_{jk},v^{(3)}_{jkl}$, $(j,k,l=1,\dots,p+q)$, such that for $n \rightarrow \infty$
\begin{align*}
\sup_\nu \abs{n^{-1}R_j^{(1)}(\nu)-r_j^{(1)}(\nu)}=o_p(1) \\
\sup_\nu \abs{n^{-1}R_{jk}^{(2)}(\nu)-r_j^{(2)}(\nu)}=o_p(1)\\
\sup_\nu \abs{n^{-1}V_{jk}^{(2)}(\nu)-v_j^{(2)}(\nu)}=o_p(1)\\
\sup_\nu \abs{n^{-1}V_{jkl}^{(3)}(\nu)-v_{jkl}^{(3)}(\nu)}=o_p(1)
\end{align*}

\item[(B)]
For $j=0,\dots,p$ and $k=0,\dots,q$, and  $n \rightarrow \infty$ 
\begin{align*}
n^{-1/2} \sup_{s,i=1,\dots,n} \abs{X_{ij}(s)}=o_p(1)\\
n^{-1/2} \sup_{s,i=1,\dots,n} \abs{Z_{ik}(s)}=o_p(1)
\end{align*}
\item[(C)]
For every $\nu$, the matrix $(r^{(2)}_{ij}(\nu))$is non-singular.
\item[(D)]  $\sup_{\norm{m}_{2}\leq 1} \norm{(I-\widetilde E)^{-1}m}_\infty< \infty$,
\item[(E)] The random variables $(a_i)_{i=1,\dots,n}$ are $iid$, independent of (X,Z) and are  absolutely continuous with continuous density $h$.
\end{enumerate}
%

\subsection{Proof of Theorem \ref{thm:asymptotics}}

We first prove the central limit theorem:
\[
n^{-1/2} (\widehat m - m) \rightarrow  U.
\]
We write
\[
n^{-1/2} (\widehat m - m )(\nu)= n^{-1/2} \int_0^\nu R(y)^{-}\begin{pmatrix}  dM(y)   \\
dM^{-a_\bullet}(y)
\end{pmatrix}= \begin{pmatrix} \mathcal M_1  \\
\mathcal M_2
\end{pmatrix} =\mathcal M.
\]
Since $\mathcal M_1$  and $\mathcal M_2$  are square integrable martingales (each with respect to its natural filtration),
$\mathcal M$ is tight  under the condition that its jumps are uniformly bounded. 
This follows from assumption (B), so $\mathcal M$ is indeed tight.
Furthermore, under assumption (A), $\mathcal M$ is asymptotically uniformly close to
\[
\overline{ \mathcal M}=n^{-1/2} \sum_i \int_0^\nu r^{(1)}(y)^{-}\begin{pmatrix}  dM_i(y)   \\
dM_i^{-a_\bullet}(y),
\end{pmatrix}
\]
which is the sum of $n$ $iid$ random processes.
So the limit of $\mathcal M$, if it exists, must be Gaussian.
Hence convergence of $\mathcal M$ to $U$ is verified by establishing point-wise convergence of the covariance matrix  of $\mathcal M$
to the covariance matrix of $U$.
For two points $\nu_1, \nu_2$ in $[0,a_{max}]$ with $\nu_1\leq\nu_2$,
$\textrm{Cov}\left(\mathcal M(\nu_1),\mathcal M(\nu_2)\right)$ is a $(p+q) \times (p+q)$ matrix.
We have
\begin{align*}
\textrm{Cov}\left(\mathcal M(\nu_1),\mathcal M(\nu_2)\right)=\begin{pmatrix} \textrm{Cov}( \mathcal M_1(\nu_1),\mathcal M_1(\nu_2))&\textrm{Cov}( \mathcal M_1(\nu_1),\mathcal M_2(\nu_2))  \\
\textrm{Cov}( \mathcal M_2(\nu_1),\mathcal M_1(\nu_2))&\textrm{Cov}( \mathcal M_2(\nu_1),\mathcal M_2(\nu_2))\end{pmatrix}.
\end{align*}
With entry $(j,k)$ given by
\begin{align*}
&\sum_{i,l,m}\textrm{Cov}\left(  \int_0^{\nu_0(\nu_1,\nu_2,a_i,j,k)} ({R^{(2)}(\nu)}^{-1})_{jl}V_{il}(\nu) dM_i(\nu), \right.\\
&\left. \qquad \qquad  \int_0^{\nu_0(\nu_1,\nu_2,a_i,j,k)} ({R^{(2)}(\nu)}^{-1})_{km}(\nu)V_{im}(\nu) \mathrm dM_i(\nu)  \right),
\end{align*}
where
\[
\nu_0(\nu_1,\nu_2,a_i,j,k)=
\begin{cases}
\nu_1 \quad \text{for  $j\leq p, k\leq p$}\\
\min(\nu_1,\nu_2-a_i) \quad \text{for}\   j \leq p, k > p\\
\nu_1-a_i \quad \text{for} \  j>p, k \leq p\\
\nu_1-a_i \quad \text{for} \ j>p,  k> p
\end{cases}.
\]
The two processes in the covariance are running in in the same time-interval.
This is because we could eliminate the non-intersecting time points due to independence.
Under assumption (B), the entries converge  to
\begin{align*}
&\sum_{i,l,m}\textrm{Cov}\left(  \int_0^{\nu_0(\nu_1,\nu_2,a_i,j,k)} ({r^{(2)}(\nu)}^{-1})_{jl}V_{il}(\nu) dM_i(\nu), \right.\\
&\left. \qquad \qquad  \int_0^{\nu_0(\nu_1,\nu_2,a_i,j,k)} ({r^{(2)}(\nu)}^{-1})_{km}(\nu)V_{im}(\nu) \mathrm dM_i(\nu)  \right),
\end{align*}
so that the two process in the covariance are now even martingales with respect to the same filtration $\mathcal F_i(\nu_0)=
\sigma\{V_i(u), N_i(u), u\leq \nu_0\}$. We can hence first calculate  the conditional covariance, given $\mathcal F_i$, i.e.,  the predictable covariation process. Afterwards, the covariance is given as the expectation of predictable covariation process.
For the predictable covariation process we get
\begin{align*}
\sum_g \int_0^{\nu_0(\nu_1,\nu_2,a_i,j,k)} ({r^{(2)}(\nu)}^{-1})_{jl}({r^{(2)}(\nu)}^{-1})_{km}V_{il}(\nu)V_{im}(\nu)V_{ig}(\nu)\begin{pmatrix}
\alpha(\nu)\\ \beta^{a_i}(\nu)
\end{pmatrix}\mathrm d\nu.
\end{align*}

From assumptions (A),(C), (E) we conclude that $\textrm{Cov}\left(\mathcal M(\nu_1),\mathcal M(\nu_2)\right) \rightarrow \Sigma(\nu_1,\nu_2)$ 
with entries
\begin{align*}
\Sigma_{jk}=\sum_{l,m,g} \int h(x) \int_0^{\nu_0(\nu_1,\nu_2,x,j,k)} ({r^{(2)}(\nu)}^{-1})_{jl}({r^{(2)}(\nu)}^{-1})_{km}v^{(3)}_{lmg}(\nu)\begin{pmatrix}
\alpha(\nu)\\ \beta^{x}(\nu)
\end{pmatrix}\mathrm d\nu \mathrm dx.
\end{align*}
Since the integral is well defined, we conclude convergence of $\mathcal M$ to $U$.

We now need to handle the operator $\overline E$.
We have
\begin{align}
&\sup_{\norm{m}_{2}\leq 1} \norm{ (\overline E - \widetilde E )m}_\infty= o_p(1), \label{1}\\
&\sup_{\norm{m}_{2}\leq 1} \norm{  \widetilde E m}_\infty< \infty. \label{2}
\end{align}
Equation \eqref{1} follows directly from the uniform convergence of the kernel functions $ E_1 (s|u), E_2 (u|s)$
to $\widetilde E_1 (s|u), \widetilde E_2 (u|s)$ which is ensured via Assumptions (A)-(C), (E).
Inequality \eqref{2} is ensured, since the kernel functions are bounded using the same assumptions.
Together with Assumption (D) it follows that the operator  $(I-\overline E)^{-1}$ converges to the linear and bounded operator $(I-\widetilde E)^{-1}$ which gives the desired central limit theorem.
\subsection{Poof of proposition \ref{prop:bootstrap}}
For two points $\nu_1, \nu_2$ in $[0,a_{max}]$ with $\nu_1\leq\nu_2$, the covariance of $\widehat {\mathcal M}^{(1)}$ is given by
\begin{align*}
&\sum_{i,l,m}\textrm{Cov}\left(  \int_0^{\nu_0(\nu_1,\nu_2,a_i,j,k)} ({R^{(2)}(\nu)}^{-1})_{jl}V_{il}(\nu) G_i dN_i(\nu), \right.\\
&\left. \qquad \qquad  \int_0^{\nu_0(\nu_1,\nu_2,a_i,j,k)} ({R^{(2)}(\nu)}^{-1})_{km}(\nu)V_{im}(\nu) G_i \mathrm dN_i(\nu)  \right).
\end{align*}
Under assumption (B) this is uniformly close to 
\begin{align*}
&\sum_{i,l,m}\textrm{Cov}\left(  \int_0^{\nu_0(\nu_1,\nu_2,a_i,j,k)} ({r^{(2)}(\nu)}^{-1})_{jl}V_{il}(\nu) G_i dN_i(\nu), \right.\\
&\left. \qquad \qquad  \int_0^{\nu_0(\nu_1,\nu_2,a_i,j,k)} ({r^{(2)}(\nu)}^{-1})_{km}(\nu)V_{im}(\nu) G_i \mathrm dN_i(\nu)  \right).
\end{align*}
As in the proof of Theorem 1, the two processes in the covariance are martingales with respect to $\mathcal F_i$,
so we can calculate the covariance as expectation of the predictable covariation process.
Hence,
$\textrm{Cov}\left(\widehat{\mathcal M}^{(1)}(\nu_1),\widehat{\mathcal M}^{(1)}(\nu_2)\right) \rightarrow \Sigma(\nu_1,\nu_2)$.

\bibliographystyle{biometrika}
\bibliography{jan20.bib}

\end{document}